%% file: 99_main.tex
\documentclass[conference]{IEEEtran}
\IEEEoverridecommandlockouts
\usepackage{amsmath,amssymb,amsfonts}
\usepackage{graphicx}
\usepackage{textcomp}
\usepackage{xcolor}
\usepackage[a4paper, 
    total={184mm,239mm}, 
    columnsep=6.3mm, 
    tmargin=19mm, 
    bmargin=43mm,
    rmargin=13mm, 
    lmargin=13mm
]{geometry}

\usepackage{pbalance}
\usepackage[nolist,nohyperlinks]{acronym}
\usepackage{breakcites} 
\usepackage{url}
\usepackage{subcaption}
\usepackage{balance}
\usepackage{layouts}

\acrodef{mcu}[MCU]{microcontroller}
\acrodef{cnn}[CNN]{convolutional neural network}
\acrodef{har}[HAR]{human activity recognition}
\acrodef{fpu}[FPU]{floating-point unit}
\acrodef{simd}[SIMD]{single instruction, multiple data}
\acrodef{mac}[MAC]{multiply-accumulate operation}
\acrodef{nas}[NAS]{neural architecture search}

\newcommand{\besttimesaving}{24\%}

\printinunitsof{in}
\newcommand\thefont{\expandafter\string\the\font} 

\newcommand{\fakepar}[1]{\vspace*{0.15cm}\noindent\textbf{#1.}}
\newcommand{\capt}[1]{\mdseries{\emph{#1}}}

\def\BibTeX{{\rm B\kern-.05em{\sc i\kern-.025em b}\kern-.08em
    T\kern-.1667em\lower.7ex\hbox{E}\kern-.125emX}}
\begin{document}

\title{Efficient CNN Inference on Ultra-Low-Power MCUs via Saturation-Aware Convolution
\thanks{This work is supported by the Swedish Foundation for Strategic Research (SSF) grant FUS21-0067.
Part of the data analysis was enabled by resources in project UPPMAX 2025/2-258 provided by Uppsala University at UPPMAX. }
}

\author{\IEEEauthorblockN{Shiming Li}
\IEEEauthorblockA{\textit{Uppsala University}\\
Sweden \\
shiming.li@it.uu.se
}
\and
\IEEEauthorblockN{Luca Mottola}
\IEEEauthorblockA{\textit{Politecnico di Milano, RISE,}\\
\textit{and Uppsala University}\\
Italy and Sweden \\
luca.mottola@polimi.it
}
\and
\IEEEauthorblockN{Yuan Yao}
\IEEEauthorblockA{\textit{Uppsala University}\\
Sweden \\
yuan.yao@it.uu.se
}
\and
\IEEEauthorblockN{Stefanos Kaxiras}
\IEEEauthorblockA{\textit{Uppsala University}\\
Sweden \\
stefanos.kaxiras@it.uu.se
}
}

\maketitle

\begin{abstract}
\input{00_abstract}
\end{abstract}

\section{Introduction}
\input{01_intro}


\section{Saturation-Aware Convolution} \label{sec:design}
\input{03_design}


\section{Evaluation}
\input{05_eval}


\section{Conclusion}
\input{07_conc}


\bibliographystyle{ieeetr}
\clearpage
\balance
\bibliography{refs}

\end{document}

%% file: 00_abstract.tex
Quantized CNN inference on ultra-low-power MCUs incurs unnecessary computations in neurons that produce saturated output values. These values are too extreme and are eventually clamped to the boundaries allowed by the neuron.
Often times, the neuron can save time by only producing a value that is extreme enough to lead to the clamped result, instead of completing the computation, yet without introducing any error. 
Based on this, we present \emph{saturation-aware convolution}: 
an inference technique whereby we alter the order of computations in convolution kernels to induce earlier saturation, and value checks are inserted to 
omit unnecessary computations when the intermediate result is sufficiently extreme. 
Our experimental results display up to \besttimesaving{} inference time saving on a Cortex-M0+ MCU, with zero impact on accuracy.

%% file: 01_intro.tex
TinyML applications such as \acp{cnn} empowers complex data processing on ultra-low-power devices under power-scarce environments~\cite{EnergyHarvesting10.1145/2915918,IoBatterylessT10.1145/3624718,IntermittentTinyMLmendis2025special}. 
The \ac{cnn} inference latency translates almost linearly to energy consumption~\cite{TADA10.1145/3666025.3699347,romano2026neuro} as these \acp{mcu} mostly lack features such as DVFS~\cite{ahmed2019betrayal,ahmed2020intermittent}. 

\fakepar{Time is wasted on neurons producing saturated values}
The majority of \acp{cnn}' inference time is spent on \acp{mac} in convolutional and fully-connected layers. 
Their operation can be generally described as $a = bias +\sum^{m-1}_{i=0}x_{i} \cdot w_{i}$,
where \( a \) is the accumulation, \( m \) is the number of computation steps, \( x_i \) are the inputs, \( w_i \) and \( bias \) represent the weights and bias, respectively. 
However, $a$ is not fed to the output neuron directly. 
It may be too large or too small, and causes the output neuron to saturate. 
Then, the actual value of the neuron is clamped within specified boundaries, which can be introduced by, for example, the value range of the neuron's data type. 
Computation time is wasted, because regardless of $a$, a saturated neuron's value is fixed.

\begin{figure}[]
    \centering
    \includegraphics[width=0.48\textwidth]{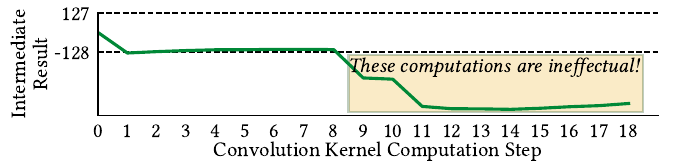}
    \caption{An example of the computation trace of a convolution operation, where the output neuron's final value saturates. }  
    \label{fig:acc_change_no_errorbar}
\end{figure}

We exemplify this in Figure~\ref{fig:acc_change_no_errorbar}, showing the computation trace of a convolution operation for an output neuron in the hand gesture recognition \ac{cnn} from the STM32 AI Model Zoo~\cite{STMAIModelZoo}. 
Since the output is of type \texttt{int8}, the result of the convolution is clamped to -128. 
The last 10 steps are wasted. 

We find that this phenomenon is common in \acp{cnn} on ultra-low-power \acp{mcu}, which are typically quantized networks where the data types have smaller value ranges and often use ReLU-like activation functions, further tightening the boundaries. 
\acp{cnn} like GMP~\cite{STMAIModelZoo} also have dynamic boundaries~\cite{PredBasedDNNExec8416870} that bear similar effects. 
Existing literature extensively explores redundancy in neural networks, for example, in weights or connections~\cite{Pruninghan2015learning,Pruninghan2015deep,barjiami24stt}, value precision~\cite{Pruninghan2015deep,QuantizationTrainingcourbariaux2014training,QuantizationTrainingJMLR:v18:16-456,QuantizationTrainingjacob2018quantization}, and even feature maps~\cite{DynamicConvolutionsVerelst_2020,DynaSpa10.1145/3666025.3699348} or network structure~\cite{BranchyNetteerapittayanon2016branchynet,Multi-scaleHuang2017MultiScaleDN,zerotimewastewolczyk2021zero,HarvNet10.1145/3581791.3596845,EnergyBasedDnnCompression10.5555/3437539.3437784,PredBasedDNNExec8416870}.
Most of these solutions fail to exploit redundancy at per-neuron granularity. 
Closest to our work is SnaPEA~\cite{Snapea8416863}. 
It analyzes  neuron-level redundancy by terminating computations that cannot produce a positive result, by relying on the condition that the inputs of each layer are all positive values. 

\fakepar{We omit computations from saturated neurons}
We propose an inference technique which executes convolution kernels while dynamically omitting the unnecessary computations in saturated neurons \emph{without introducing error}. 
We integrate our design into a modified, arm-v6m compatible version of TinyEngine, the code generation and inference engine of the state-of-the-art MCUNet~\cite{MCUNetlin2020mcunet}. 
We conduct experiments on a STM32 development board with a Cortex-M0+ \ac{mcu} and observe up to \besttimesaving{} time saving across 7 \acp{cnn} we test. 

%% file: 03_design.tex
\begin{figure}[]
  \centering
    \begin{minipage}[b][3cm][s]{0.175\textwidth}
        \begin{subfigure}[]{1\textwidth}
            \centering
            \includegraphics[width=1\textwidth]{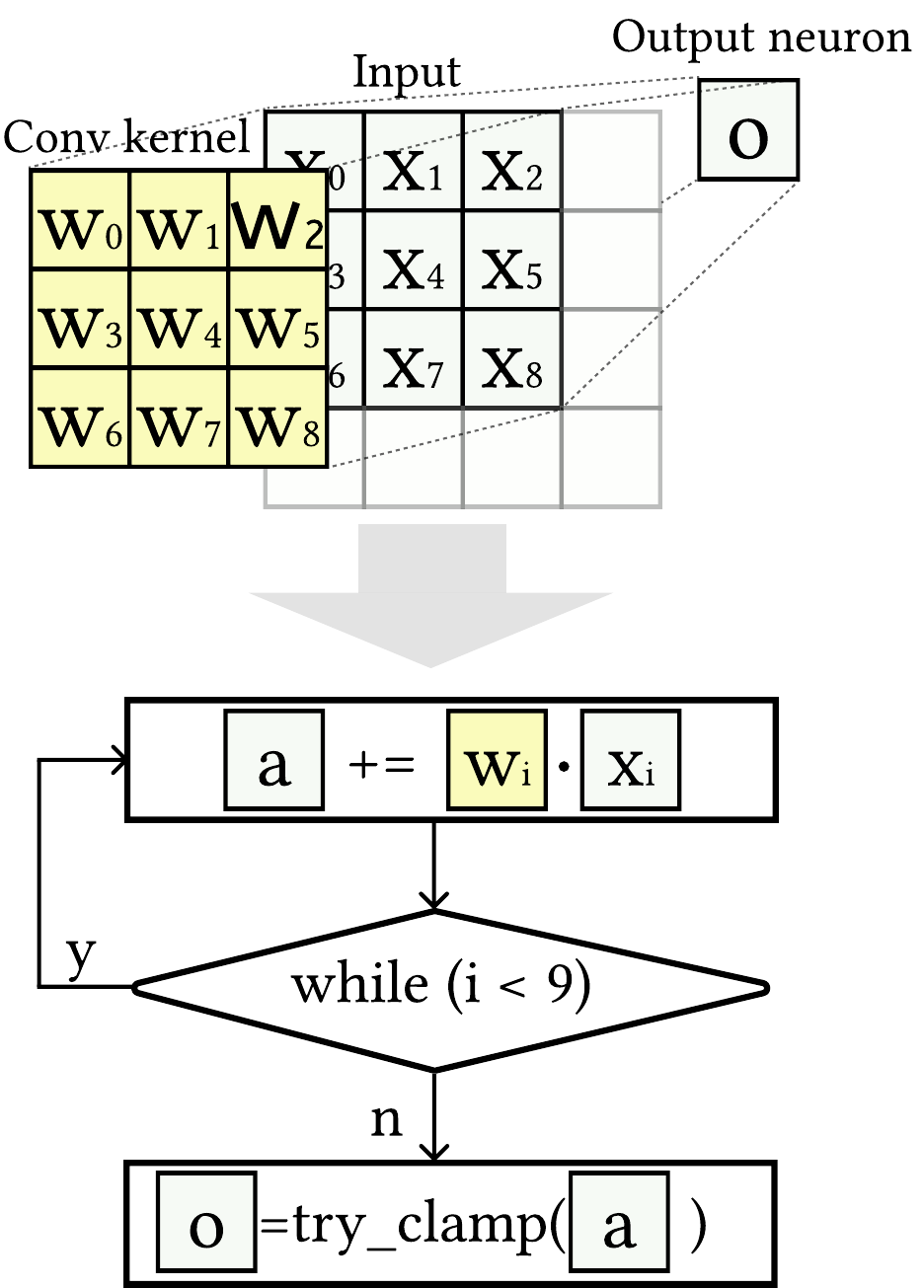}
            \caption{\capt{Conventional convolution operation.}}
            \label{fig:conventional_conv}
         \end{subfigure}
    \end{minipage} 
    \begin{minipage}[b][3cm][s]{0.30\textwidth}
        \begin{subfigure}[]{1\textwidth}
            \centering
            \includegraphics[width=1\textwidth]{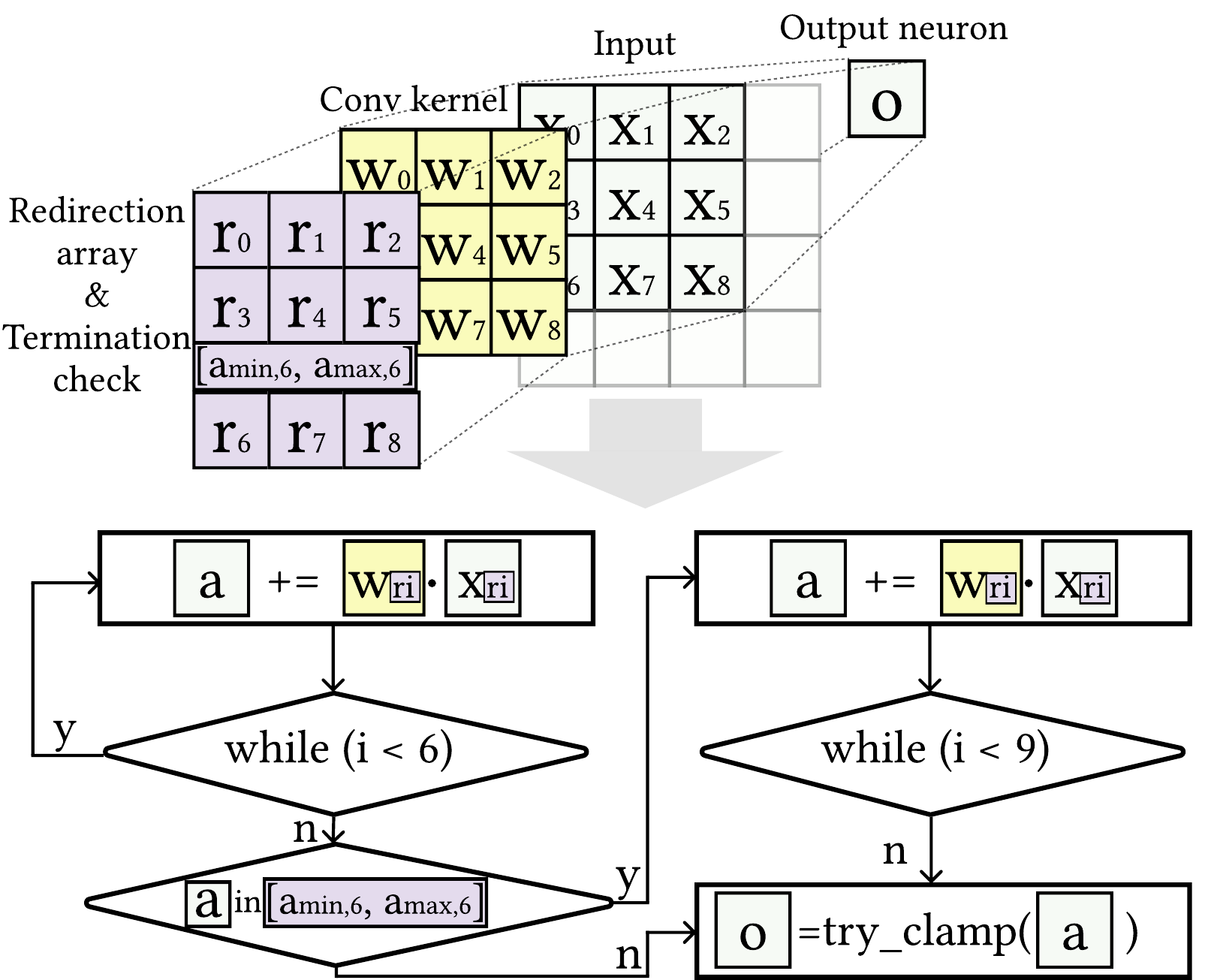}
            \caption{\capt{Saturation-aware convolution. }}
            \label{fig:saturation_aware_conv}
        \end{subfigure} 
    \end{minipage}
    \caption{Conventional and saturation-aware convolution. }
    \label{fig:compare_conv}
\end{figure}

We propose saturation-aware convolution, where convolution kernels allow neurons that would definitely produce a saturated value to terminate computation early, without error. 

\fakepar{Omitting ineffectual computations with zero error} 
We only allow terminating a convolution operation when the intermediate result is too extreme, and is bound to cause saturation. 
This can be done for any computation step by checking the most extreme possible future result. 
For example, in the case of an \texttt{int8} \ac{cnn}, any future input value $x$ is in $[-128, 127]$. 
Already knowing the weight value $w_{i+1}$ at computation step $i+1$, the result of this step must be within $[-128 \cdot w_{i+1}, 127 \cdot w_{i+1}]$ if $w_{i+1}$ is non-negative, or $[127 \cdot w_{i+1}, -128 \cdot w_{i+1}]$ if $w_{i+1}$ is negative. 
Summing up all possible limits of future results, we know $a_i$'s future deviation will not exceed $[d_{min,i},d_{max,i}]$, where $
    d_{min,i} = \sum^{m-1}_{j=i+1}
    \scriptsize\begin{cases}
    -128 \cdot w_j, \text{if }w_j \geq 0\\
    127 \cdot w_j, \text{if }w_j < 0
    \end{cases}
$, 
and $
    d_{max,i} = \sum^{m-1}_{j=i+1}
    \scriptsize\begin{cases}
    127 \cdot w_j, \text{if }w_j \geq 0\\
    -128 \cdot w_j, \text{if }w_j < 0
    \end{cases}
$. 

Note that values involved to compute $d_{min,i}$ and $d_{max,i}$ are compile-time knowledge, easing the need for extra computation at run-time. 
A convolution kernel can practically utilize the information of $d_{min,i}$ and $d_{max,i}$ to compare $a_i$ against a range $[a_{min,i},a_{max,i}]$, where $a_{min,i}=-128-d_{max,i}$ and $a_{max,i}=127-d_{min,i}$. If $a_i$ has left the range, it means even the most extreme future deviation of $a$ cannot bring it back into $[-128, 127]$, and $a_i$ can be deemed ``too extreme'' and will definitely lead to saturation.

\fakepar{Reordering computations to induce earlier saturation}
The computation order in a convolution kernel affects the changes of intermediate results. 
Intuitively, we prefer the intermediate value of $a$ to stabilize more quickly ~\cite{ApproximateIntermittentbambusi2022case}, leaving less space for deviation in the future. 
Thus, we reorder the computation in a convolution kernel by the absolute value of weight $w_i$. 

The order does not affect the final result of $a$, since in a quantized \acp{cnn} with integer types, the additions and multiplications are commutative. 
However, it effectively increases the percentage of unnecessary computations that can be omitted without introducing error. According to our experiments, this is increased by four times, from 5\% to 20\%, as discussed next.

\fakepar{Saturation-aware convolution execution flow}
We illustrate and compare conventional and saturation-aware convolution in Figure~\ref{fig:compare_conv}. 
The output neuron's value $o$ is produced by filtering out any extreme value of $a$ through clamping. 
The example saturation-aware convolution operation in Figure~\ref{fig:saturation_aware_conv} executes convolution with extra information. 
In this example, a termination check is inserted at the sixth computation step. 
The check examines the intermediate result of $a$ against $[a_{min,6},a_{max,6}]$, and if the value is too extreme, the convolution operation will be terminated, and the intermediate value of $a_6$ will be used to produce the result to be fed to the neuron. 
Moreover, instead of executing $x_i \cdot w_i$ in the natural number order, Figure~\ref{fig:saturation_aware_conv} introduces a redirection array to induce earlier saturation.

%% file: 05_eval.tex
We implement saturation-aware convolution and evaluate \ac{cnn} model inference speed and energy consumption. 
Experiments are conducted on the STM32 NUCLEO-G0B1RE development board with a Cortex-M0+ MCU, using 7 lightweight \acp{cnn} from the open-source STM32 Model Zoo~\cite{STMAIModelZoo,STMAIModelZooServices}. 

\fakepar{Implementation}
We implement scripts to analyze models, generate the computation order and find the layers where to apply saturation-aware convolution. 
We allow up to 2 checks per kernel, and we insert them in each convolution kernel at the positions where the most computation steps can be omitted statistically, which we decide by profiling the kernels' behavior with sample inputs not used in the evaluation. 

We integrate saturation-aware convolution into a modified, arm-v6m compatible version of TinyEngine~\cite{MCUNetlin2020mcunet}, the state-of-the-art neural network code generation and inference engine for \acp{mcu}. 
TinyEngine takes in \ac{cnn} models in TenforFlow Lite \verb|.tflite| format, and generates \verb|.c| code that can be compiled and run on \acp{mcu}.

\begin{figure}[tb]
    \centering
    \includegraphics[width=0.48\textwidth]{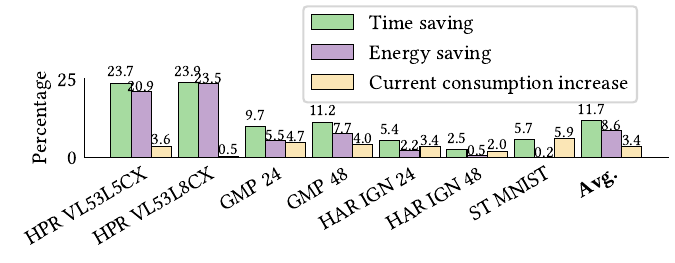}
    \caption{Inference latency reduction, current consumption increase and energy saving with saturation-aware convolution.}
    \label{fig:savings}
\end{figure}

\fakepar{Results}
Saturation-aware convolution effectively reduces inference latency and gains energy saving. 
We display the time and energy saving in Figure~\ref{fig:savings}. 
We achieve \textit{up to 23.9\%} time saving, and up to 23.5\% energy saving on HPR VL53L8CX, averaging 11.7\% and 8.6\%, respectively, across all workloads.  
We observe only a small increase in current consumption with saturation-aware convolution, only 3.6\% on average, as in Figure~\ref{fig:savings} (yellow bar).
We compare the \acp{cnn}'s outputs for each single experiment with the baseline, and verify that our technique introduces \emph{strictly zero error}.


%% file: 07_conc.tex
We present saturation-aware convolution, where a convolution kernel executes the computations in an altered order to induce saturation and predicts saturation dynamically via infused compile-time information, without introducing error. 
Based on experiments involving 7 \ac{cnn} workloads, our technique shows up to 24\% inference latency reduction and up to 23.5\% energy saving, with zero impact on accuracy.